\documentclass[english,reviewcopy]{elsarticle}
\usepackage[T1]{fontenc}
\usepackage[latin9]{inputenc}
\usepackage[a4paper]{geometry}
\usepackage{amsmath}
\usepackage{graphicx}
\usepackage{setspace}
\usepackage{amssymb}
\makeatletter
\@ifundefined{definecolor}
 {\@ifundefined{definecolor}
 {\@ifundefined{definecolor}
 {\usepackage{color}}{}
}{}
}{}
\definecolor{labelcolor}{RGB}{100,0,0}
\makeatother

\makeatother

\usepackage{babel}

\makeatother

\usepackage{babel}

\begin{document}

\title{Preservation of entanglement in a two-qubit-spin coupled system}
\author{Yu-Chen Hou}
\address{School of Electronic and Information Engineering, Beihang University,
Xueyuan Road No. 37, Beijing 100191, PR China}
\author{Guo-Feng Zhang}
\ead{Corresponding author: gf1978zhang@buaa.edu.cn}
\address{School of Physics and Nuclear Energy Engineering, Beihang University,
Xueyuan Road No. 37, Beijing 100191, PR China}
\author{Yan Chen}
\address{School of Physics and Nuclear Energy Engineering, Beihang University,
Xueyuan Road No. 37, Beijing 100191, PR China}
\author{Heng Fan}
\address{Beijing National Laboratory for Condensed Matter Physics, Institute of Physics, Chinese Academy of Sciences, Beijing 100190, PR China}

\begin{abstract}
A theoretical scheme to preserve the entanglement in a two-qubit-spin
coupled system in the presence of Dzyaloshinskii-Moriya (DM) anisotropic
antisymmetric interaction is proposed. Based on a sequence of operations
performed periodically on the system, the scheme can preserve the
entanglement of the system starting from any initial state with any
high precision for any long duration.

PACS:03. 67. Lx, 03. 65. Ta
\end{abstract}

\maketitle

\section{Introduction}

Entanglement plays a central role in quantum information processing.
However, entanglement is so fragile that in real circumstances it
cannot be easily preserved. Therefore, the implementation of preservation
of entanglement is a crucial element in the realization of quantum
information processing. Last decade has seen a number of theories,
and techniques on the preservation of entanglement, most of which
are based on formalized mathematics models and quantum optical systems. Some
examples are error-avoiding \cite{cpy}, bang-bang control \cite{lvi} and Super-Zeno effect \cite{ddh}. The well-known Quantum
Zeno effect, which was first presented in the literature as a paradoxical consequence of measurements on quantum mechanics \cite{bmi}, is also
a useful tool for quantum state protection \cite{pfa}, entanglement control \cite{jgo} and entanglement preservation \cite{sma}. But quantum optical systems,
like their classical counterparts, have an intrinsic disadvantage:
poor scalability. Among other schemes that have been proposed to construct
quantum information systems, the ones based on solid state systems
are believed to have the best scalability \citep{jlo}. Moreover,
the solid state schemes can largely take advantage of modern semiconductor
technology and micro-fabrication technology. Because of these advantages,
the study of solid state quantum computation is currently an important
field \citep{dlo,bek,gbu,lfs,jle}. Solid spin systems are among those
solid-based candidates for the realization of the entanglement. The
spin chains not only have useful applications such as the quantum
state transfer, but also display rich entanglement features \citep{sbo,mch}.
The Heisenberg chain, the simplest spin chain, has been used to construct
a quantum computer and quantum dots \citep{dlo}. By suitable coding,
the Heisenberg interaction alone can be used for quantum computation
\citep{lfs,dal,dpd}. Obviously, researches on the preservation of
entanglement in spin systems will provide a possibility of an ideal
physical foundation for quantum information systems. The objective
of the present work is to construct a scheme to control the evolution
of a two-qubit-spin system (a typical case of a general spin system)
and preserve entanglement. In this paper, the scheme is described
first and then an example is given to illustrate the effect of the
scheme.

\section{Model and scheme}

In this paper, we consider the Heisenberg model with DM interaction
which can be written as \begin{equation}
H=J_{1}\sigma_{1x}\sigma_{2x}+J_{2}\sigma_{1y}\sigma_{2y}+D(\sigma_{1x}\sigma_{2y}-\sigma_{1y}\sigma_{2x}),\end{equation}
 where $\sigma_{mn}\left(m=1,2;n=x,y\right)$'s are Pauli matrices,
$J_{i}(i=1,2)$'s are the real coupling coefficients and $D$ is the
DM vector coupling which is assumed only along $z$ axis. The DM anisotropic
antisymmetric interaction arises from spin-orbit coupling \citep{idz,tmo1,tmo2,tmo3}.
We can control the system by interrupting the free evolution periodically
with a certain operation, so that the state of the controlled evolution
is confined in a subset of a state space very close to the initial
state and the entanglement can be preserved very well. This idea is
inspired by prior work of R. Rossi \citep{rro}.

The free evolution $U_{f}$ governed by the Hamiltonian is: \begin{equation}
U_{f}\left(t\right)=e^{-iHt/\hbar}.\end{equation}

In order to preserve the initial entanglement , we can interrupt the
free evolution before the state deviates from the initial state too
much, and guide the state back to its starting point. We call this
process the \emph{cyclic }evolution. The implementation of such a
process is given as follows.

The cyclic evolution $U_{cT}$ is constructed in this way: the system
evolves freely for some time $T$, then a certain operation $O$ is
applied to the system, then a same evolution and a same operation,
which can be written as \begin{equation}
U_{cT}\left(\tau\right)=\begin{cases}
U_{f}\left(0\right)\equiv I & \tau=0\\
U_{f}\left(\tau\right) & 0<\tau<T\\
U_{f}\left(T\right) & \tau\rightarrow T_{-}\\
OU_{f}\left(T\right) & \tau\rightarrow T_{+}\\
U_{f}\left(\tau-T\right)OU_{f}\left(T\right) & T<\tau<2T\\
U_{f}\left(T\right)OU_{f}\left(T\right) & \tau\rightarrow\left(2T\right)_{-}\\
OU_{f}\left(T\right)OU_{f}\left(T\right)\equiv I & \tau\rightarrow\left(2T\right)_{+}\end{cases},\label{eq:cyclic}\end{equation}
 where $O=I_{1}\otimes\sigma_{2z}$. Note that the cyclic evolution
is discontinuous for two points, $\tau=T$ and $\tau=2T$, where operations
are applied. These discontinuities are demonstrated explicitly above
by one-sided limits with - and + signs indexing the below (left) limit
and above (right) limit respectively. Pay special attention to the
equation $OU_{f}\left(T\right)OU_{f}(T)\equiv I$ $\left(\forall T\right)$,
which can be proved by simple calculations. This equation shows that
the evolution is indeed a cyclic evolution (with period $2T$), at
the end of that the state is guided back to the starting point. Note
the graduality of the free evolution at the initial state: \begin{equation}
\lim_{T\rightarrow0}U_{f}\left(T\right)=I,\end{equation}
 which suggests that the state will not deviate from the starting
point very fast. So we can expect to preserve any initial state to
any desired precision in such a cycle, by shortening the time interval
$T$ to a sufficient level. This will be presented explicitly in Eq.(11)
and Fig.2.

High precision of the preservation requires the time interval $T$
to be short. Meanwhile, we want the preservation lasts as long as
possible. We can manage to achieve both of these two points (high
precision and long duration) by fully taking advantage of the \emph{cyclic
}character of the cyclic evolution. In fact, the controlled evolution
is implemented by repeating the same cyclic evolution (interrupting
the free evolution with the same operation in period $2T$): \begin{eqnarray}
 &  & U_{c}\left(t\right)\nonumber \\
 & = & U_{cT}\left(\tau\right)OU_{f}(T)OU_{f}(T)\cdots OU_{f}(T)OU_{f}\left(T\right)\nonumber \\
 & = & U_{cT}\left(\tau\right)I\cdots I\nonumber \\
 & = & U_{cT}\left(\tau\right),\end{eqnarray}
 where $\tau=t-2kT$ and $k=\left\lfloor \frac{t}{2T}\right\rfloor $is
the number of complete cycles ($\lfloor\rfloor$ is the floor operator)
so that $t=\tau+2kT$ and $0\leq\tau<2T$. In this way, the preservation
can last forever, theoretically.

The construction of the scheme to control the evolution is now completed.
Notably, there are no restrictions on the precision, the duration
or the initial state.

\section{Example}

To illustrate the effect of the scheme, we will compare the entanglement
of the free evolution and that of the controlled one in an example,
where the initial state is assumed to be in a Bell state: $|\psi\left(0\right)\rangle=\frac{1}{\sqrt{2}}|01\rangle+\frac{1}{\sqrt{2}}|10\rangle$
and the entanglement is calculated by concurrence\citep{shi}: \begin{equation}
C\left(t\right)=\max\left(0,2\max\left\{ \sqrt{\mu_{i}}\right\} -\sum_{i}\sqrt{\mu_{i}}\right),\end{equation}
 where $\mu_{i}$'s are eigenvalues of the operator $\rho S\rho^{*}S$
with $S=\sigma_{1y}\otimes\sigma_{2y}$ and $\rho=U|\psi\left(0\right)\rangle\langle\psi\left(0\right)|U^{\dagger}$.

Applying Eq.(2), Eq.(3) and Eq.(5) to equation Eq.(6), we obtain the
concurrences of the free, cyclic and controlled evolutions and their
minimums:\begin{equation}
C_{f}(t)=\frac{\sqrt{2\,{D}^{2}\,\mathrm{cos}\left(4\, t\,\sqrt{{K}^{2}+4\,{D}^{2}}\right)+{K}^{2}+2\,{D}^{2}}}{\sqrt{{K}^{2}+4\,{D}^{2}}},\end{equation}
 \begin{equation}
C_{c}\left(t\right)=C_{cT}\left(\tau\right)=\begin{cases}
\frac{\sqrt{2\,{D}^{2}\,\mathrm{cos}\left(4\,\tau\,\sqrt{{K}^{2}+4\,{D}^{2}}\right)+{K}^{2}+2\,{D}^{2}}}{\sqrt{{K}^{2}+4\,{D}^{2}}} & 0\leq\tau\leq T\\
\frac{\sqrt{2\,{D}^{2}\,\mathrm{cos}\left(\sqrt{{K}^{2}+4\,{D}^{2}}\,\left(8\, T-4\,\tau\right)\right)+{K}^{2}+2\,{D}^{2}}}{\sqrt{{K}^{2}+4\,{D}^{2}}} & T\leq\tau\leq2T\end{cases},\end{equation}
 \begin{equation}
\left(C_{f}\left(t\right)\right)_{\min}=\frac{\left|K\right|}{\sqrt{{K}^{2}+4\,{D}^{2}}},\end{equation}
 \begin{equation}
\left(C_{c}\left(t\right)\right)_{\min}=\left(C_{cT}\left(\tau\right)\right)_{\min}=\frac{\sqrt{2\,{D}^{2}\,\mathrm{cos}\left(4\,\sqrt{{K}^{2}+4\,{D}^{2}}\, T\right)+{K}^{2}+2\,{D}^{2}}}{\sqrt{{K}^{2}+4\,{D}^{2}}},\end{equation}
 where $K=J_{1}+J_{2}$, $\tau=t-2kT$ and $k=\left\lfloor \frac{t}{2T}\right\rfloor $
is the number of complete cycles. Note the precision of the preservation
that can be achieved: \begin{equation}
\lim_{T\rightarrow0}\left(\left(C_{c}\left(t\right)\right)_{\min}\right)=1,\end{equation}
 which indicates that the initial entanglement is preserved well by
shortening the time interval $T$, as we have expected in Eq.(4).

The effect of the scheme can be illustrated better by Fig.1, where
a comparison of concurrences of the free and the controlled evolution
is given. The concurrence of free evolution vibrates severely, while
that of the controlled one is kept at a high level close to the initial
concurrence. The controlled evolution is composed of a sequence of
cyclic evolutions. Every cyclic evolution has two stages: the first
one is a \emph{falling} stage and the second one is a \emph{rising
}stage. At the end of each stage, an operation is performed: the first
operation terminates the falling stage and initializes the rising
stage, while the second operation restores the initial state and
ensures a smooth transition from the current cyclic evolution to the
next one. Therefore, the
concurrence is kept from deviating too far away by a sequence of periodic
interruptions. %
\begin{figure}[h]
 \includegraphics[width=0.95\textwidth]{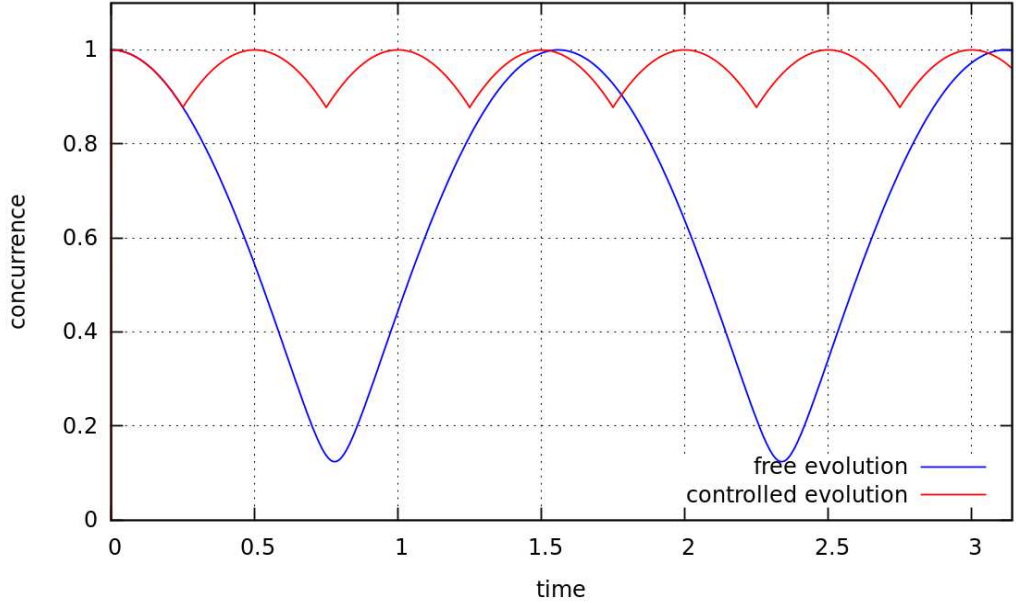}\caption{(color online) Comparison of concurrences with parameters $T=1/4$,
$D=1/2$ and $K=1/8$.}

\end{figure}

As is mentioned above in Eq.(4) and Eq.(11), the requirement of high
precision preservation can be satisfied by increasing the frequency
(shortening the time interval $T$) of the sequence of interruptions,
which is illustrated in Fig.2. %
\begin{figure}[h]
 \includegraphics[width=0.95\textwidth]{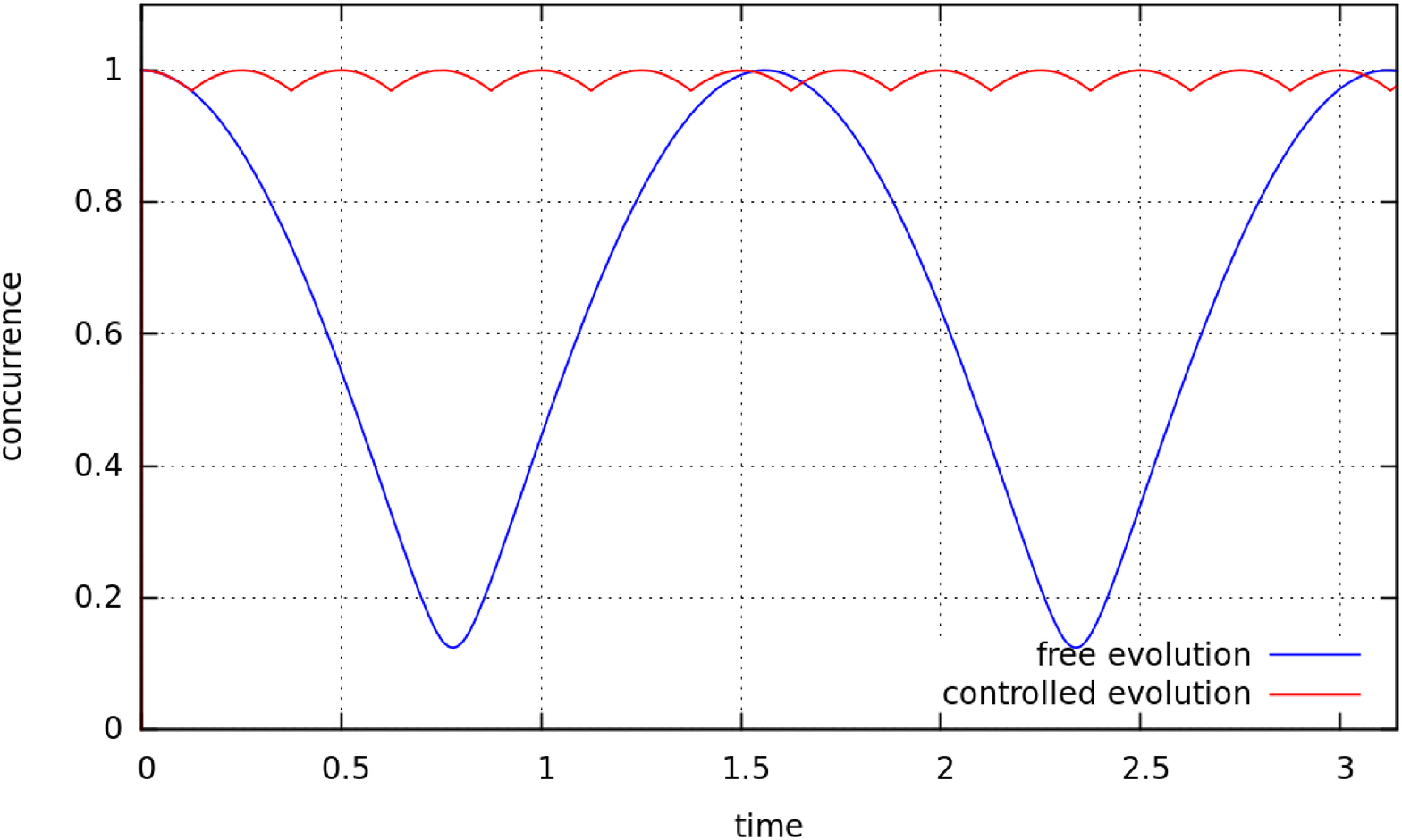}\caption{(color online) Comparison of concurrences (higher operation frequency)
with parameters $T=1/8$, $D=1/2$ and $K=1/8$.}

\end{figure}
\section{Conclusions}

We have proposed a theoretical scheme to preserve the entanglement
in a two-qubit-spin system with DM interaction by applying a sequence
of operations periodically on the system. This scheme can preserve
the entanglement of the system starting from any initial state with
any high precision for any long duration.

\section{Acknowledgements}

This work was supported by the National Science Foundation of China
under Grants No. 10874013. Heng Fan  acknowledges the support of the National Science Foundation of China under Grant No. 10974247. Thanks to the folks at Beihang
University: Shisheng Zhang at school of physics provided revising
assist; Yanlong Hou at school of electric engineering and Hua Tang
at school of electronic engineering provided software suggestions.

\end{document}